\documentclass[10pt]{article}
\usepackage[utf8]{inputenc}
\usepackage{lipsum}
\usepackage{geometry}
\usepackage{authblk}
\usepackage{verbatim}
\usepackage{amsmath}
\usepackage{amssymb}
\usepackage{amsfonts}
\usepackage{array, etoolbox}
\usepackage{longtable}
\usepackage{multirow}
\usepackage{multicol}
\usepackage{caption}
\usepackage{subcaption}
\usepackage{adjustbox}
\usepackage[dvipsnames]{xcolor}
\usepackage{hyperref}
\usepackage{url}
\usepackage{accents}
\usepackage{dsfont}
\usepackage{bbm}
\usepackage{subfiles}
\usepackage{pdflscape}

\hypersetup{
    colorlinks=true,
    linkcolor=blue,
    filecolor=magenta,      
    urlcolor=cyan,
}
\newcolumntype{P}[1]{>{\centering\arraybackslash}p{#1}}
\preto\tabular{\setcounter{magicrownumbers}{-1}}
\newcounter{magicrownumbers}

\graphicspath{{images/}{../images/}}
\title{\bf Problems with information theoretic approaches to causal learning}
\author[ ]{Nithin Nagaraj \\ Consciousness Studies Programme \\ Indian Institute of Science Campus \\ National Institute of Advanced Studies, Bengaluru, India \\ Email: nithin@nias.res.in}

\date{}

\begin{document}

\maketitle

\begin{abstract}
The language of information theory is favored in both causal reasoning and machine learning frameworks. But, is there a better language than this? In this study, we demonstrate the pitfalls of infotheoretic estimation using first order statistics on (short) sequences for causal learning. We recommend the use of data compression based approaches for causality testing since these make very little assumptions on data as opposed to infotheoretic measures, and are more robust to finite data length effects. We conclude with a discussion on the challenges posed in modeling the effects of conditioning process $X$ with another process $Y$ in causal machine learning. Specifically, conditioning can increase {\it confusion} which can be difficult to model by classical information theory. A conscious causal agent creates new choices, decisions and meaning which poses huge challenges for AI. 
\end{abstract}


\section{Introduction}

Shannon's 1948 masterpiece on a mathematical theory of communication~\cite{shannon1948} began a new era -- the Information Age which has seen unprecedented developments in the last few decades. Information theory finds important applications in artificial intelligence (AI), artificial neural networks (ANN), machine learning (ML), deep learning (DL), causality testing and the list goes on and on. Cross entropy, a closely related cousing of Shannon entropy is used to define a loss function for logistic regression. Information gain ratio~\cite{quinlan1986induction} is used in decision tree learning to overcome the limitation of information gain by reducing bias towards multi-attributes (by accounting for number and size of branches while choosing an attribute). Tishby et al.~\cite{tishby2000information} proposed the information bottlelneck method for distributional clustering and dimension reduction, and recently suggested it as a theoretical foundation for deep learning.

Causality testing is rife with rigrous application of information theory and use of infotheoretic quantities.  For example, Transfer Entropy~\cite{schreiber2000measuring}, conditional mutual information for detecting ``direction of information flow'' between coupled systems~\cite{paluvs2001synchronization}, Baghli's nonparametric characterization of causality founded upon  information-
theoretic statistics~\cite{baghli2006model}, information flow~\cite{san2016information}  and several others (see~\cite{hlavavckova2007causality} and references therein) employ infotheoretic quantities for causal inference in bivariate/multi-variate systems.

The starting point of a infotheoretic approach in all of the aforementioned areas is of course the celebrated formula of Shannon's entropy function $H(\cdot)$. For an independent and identically distributed 
(i.i.d) source $Z$, with its time series $Z_1, Z_2, \ldots, Z_n$ (also i.i.d) has the entropy $H(Z)$ given by:
\begin{equation}
    H(Z) = -\sum_{i=1}^{K} p_i \log_2(p_i), 
    \label{eqn_entropy}
\end{equation}
where $Z$ takes values from a discrete alphabet of size $K$ distinct symbols with respective probabilties $p_1$, $p_2$, $\ldots$, $p_K$ (assuming that none of these are zero). $H(Z)$ is measured in bits. Definitions of joint entropy, conditional entropy, mutual information -- all defined for two or more discrete random variables are naturally built upon $H(\cdot)$ (see~\cite{cover1999elements}, also Appendix A.1).

The connection between entropy and data compression is very tight - as proved by Shannon in the now famous source coding theorem~\cite{shannon1948, cover1999elements}. Essentially, performance of all lossless data compression algorithms is  bounded by entropy. This also means that entropy can be calculated by using lossless data compression algorithms or by measures derived from these - such as Lempel-Ziv complexity~\cite{lzcomplexity}. For example, Lempel-Ziv complexity has been employed to estimate entropy rate of spike trains~\cite{amigo2004estimating}. However, directly estimating empirical probabilities to compute the value of entropy given in Eq.~\ref{eqn_entropy} is still popular and widely used.

The innocent looking Eq.~\ref{eqn_entropy} is typically computed for a time series by first estimating the {\it first order} probabilities of the $K$ outcomes of $Z$.\footnote{We assume real-valued time series with finite length. The time series is first converted into a sequence of $K$ distinct symbols by quantization before computing entropy.} However, this is where things can get tricky as we shall demonstrate with a cleverly designed example. The rest of the paper is organized as follows -- in section 2, the pitfalls with mutual information estimation is demonstrated with an example, and how data compression based measures are immune to them. Section 3 deals with casual inference on the same example using Transfer Entropy -- one of the most popular infotheory based causality testing algorithm. A discussion on the problems with information theory in trying to model the decision making process of a conscious causal agent is explored in section 4. We conclude with Shannon's warning in section 5.

\section{Mutual information vs. data compression-based measure}
\begin{figure}[!h]
  \centering
 \includegraphics[width=0.8\linewidth]{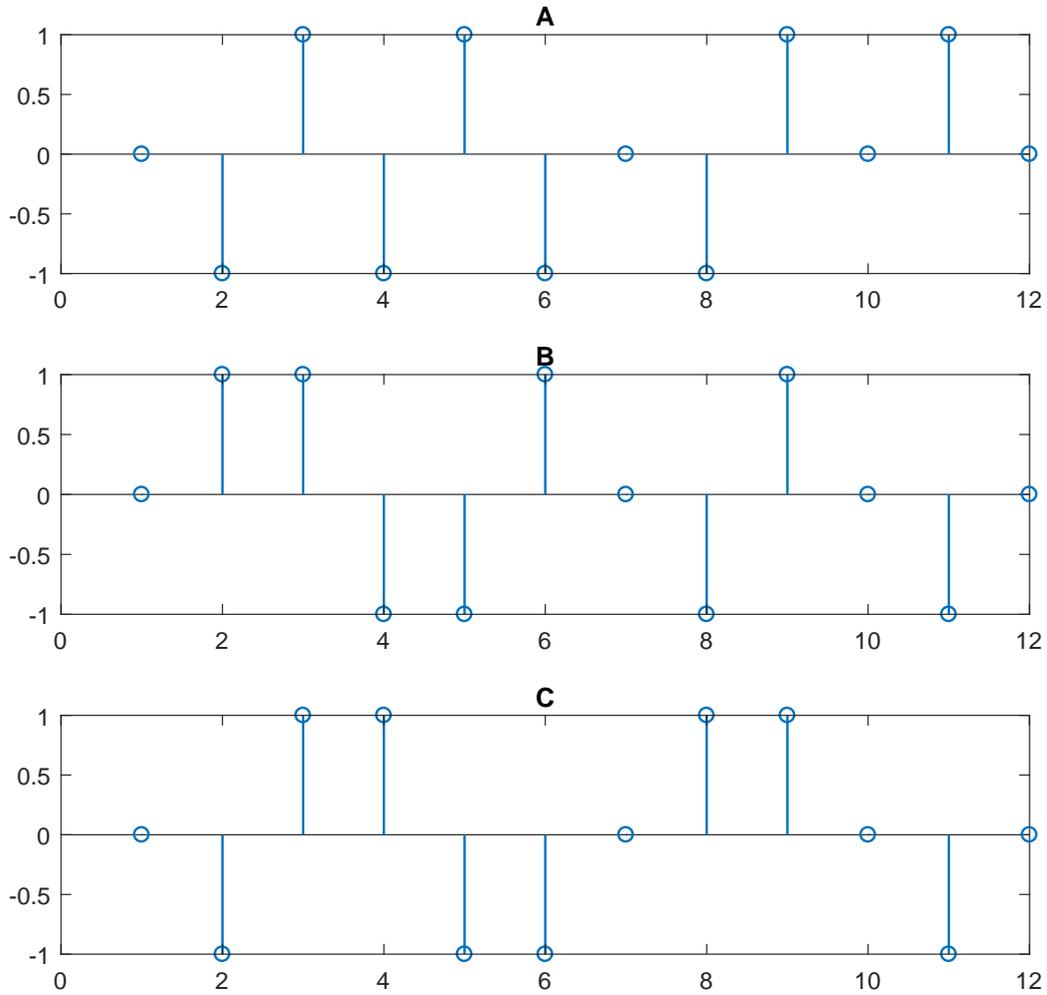}
  \caption{Three discrete-time firing patterns $A$, $B$ and $C$ of duration 12 time units. The values of firing are from the set $\{+1, -1, 0 \}$ corresponding to $\{${\bf excitatory}, {\bf inhibitory}, {\bf inactive/OFF}$\}$. All 3 take the value $0$ ({\bf inactive/OFF}) at the same time instants governed by a higher order mechanism $Z$.}
  \label{fig1}
\end{figure}

Consider three hypothetical neurons $X_1$, $X_2$ and $X_3$ which have the following firing patterns:
\begin{eqnarray*}
X_1 &=& [A A A A],\\
X_2 &=& [B C B C],\\
X_3 &=& [C B C B],
\end{eqnarray*}
where $A$, $B$ and $C$ are 12-length firing patterns shown in Fig.~\ref{fig1}.  Note that at every time instant, the hypothetical neurons can be in only one of three states:  {\bf excitatory} ($+1$), {\bf inhibitory} ($-1$) or {\bf inactive/OFF} ($0$). At any instant of time, all three neurons are simultaneously made {\bf inactive/OFF} by a higher order mechanism $Z$. Thus, the time stamps of 0s in the firing pattern coincide for all the three neurons making them (nonlinearly) correlated. $X_1$ is independent of $X_2$ and $X_3$, while $X_2$ and $X_3$ are tightly coupled such that their firing patterns are cyclic permutations of each other. 

\begin{table}[!h]
  \caption{Z is switched ON. Linear correlation coefficient $\rho$, Mutual Information (MI) and Mutual Effort-To-Compress (METC) for the multivariate system $\{ X_1, X_2, X_3 \}$.}
  \label{table1}
  \centering
  \begin{tabular}{|c|c|c|c|}
  \hline
        Pair     & $\rho$    & MI (bits) & METC \\
    \hline
    $(X_1,X_2)$ & 0 & 0.9183 & 0.1277 \\
    $(X_1,X_3)$ & 0 & 0.9183 & 0.1277 \\ 
    $(X_2,X_3)$ & 0 & 0.9183 & {\bf 0.2766} \\
    \hline
  \end{tabular}
\end{table}
In Table~\ref{table1}, we provide the pair-wise estimates of Pearson's linear correlation coefficient ($\rho$), first-order mutual information (MI, see Appendix A.1) and mutual Effort-To-Compress (METC). The Effort-To-Compress (ETC) and METC are described in the Appendix (A.2). Notice that the values of $\rho$ are zero for all pairs indicating that the firings are linearly uncorrelated. The first-order MI estimates are identical ($=0.9183$ bits) contradicting the fact that $X_2$ and $X_3$ are  highly dependent on each other than on $X_1$. But, first-order MI is unable to capture this. The data-compression based measure METC gives a value of $0.2766$ between $X_2$ and $X_3$ which is more than twice of the value for the other pairs. Thus, METC is able to better capture the tight dependence between $X_2$ and $X_3$. 
\begin{table}[!h]
  \caption{Z is switched OFF. Linear correlation coefficient $\rho$, Mutual Information (MI) and Mutual Effort-To-Compress (METC) for the multivariate system $\{ X^{'}_1, X_2^{'}, X_3^{'} \}$.}
  \label{table2}
  \centering
    \begin{tabular}{|c|c|c|c|}
  \hline
    Pair     & $\rho$    & MI (bits) & METC \\
  \hline
    $(X_1^{'},X_2^{'})$ & 0 & 0 & 0 \\
    $(X_1^{'},X_3^{'})$ & 0 & 0 & 0 \\ 
    $(X_2^{'},X_3^{'})$ & 0 & 0 & {\bf 0.0968}\\
\hline
  \end{tabular}
\end{table}

Consider the scenario where the higher order mechanism $Z$ is such that it never allows $X_1$, $X_2$ and $X_3$ to be {\bf inactive/OFF} for any time instant.  This implies that we have to remove all the time instants in the firing of the three neurons corresponding to the value $0$ to obtain the new firing patterns $X_1^{'}$, $X_2^{'}$ and $X_3^{'}$. The values of $\rho$, MI and METC are now updated as given in Table~\ref{table2}. It turns out that now the first order MI is patently wrong since it is indicating that all pairs are independent when in fact $(X_2^{'}, X_3^{'})$ continue to be cyclic permutations of each other and thus highly mutually dependent. METC is able to correctly capture all the pairwise relationships in this scenario - yielding a value of $0.0968$ for the pair $(X_2^{'}, X_3^{'})$ and zero for others. 

The aforementioned example is just a simple demonstration of the limitations of first-order MI and the power of data compression based measure (such as METC) to robustly capture the non-linear dependencies between short-length sequences. In practical applications in causality testing and machine learning, such problems are bound to occur and one must exercise utmost caution with first order estimates of infotheoretic quantities (MI in this instance, but it could be others such as entropy, conditional entropy, cross entropy, relative entropy etc.). ETC, METC and similar data compression inspired measures don't make explicit assumptions about stationarity or ergodicity. They work by parsing the given sequence and recognizing and learning patterns.

\section{Causal inference using Transfer Entropy (TE)}
In order to explicitly demonstrate the problems in causal reasoning using infotheoretic ideas, we employ a popular and widely used causal discovery algorithm known as Transfer Entropy (TE)~\cite{schreiber2000measuring, bossomaierintroduction}.
TE has been applied in a large number of applications~\cite{bossomaierintroduction}, including estimation of causality in multivariate systems.
In order to estimate the causal influence of a time series $x$ on $y$, the idea behind TE is to estimate the Kullback-Leibler (KL) divergence between the two  distributions $p(x_{n+1}| x_n^{(k)}, y_n^{(l)})$ and 
$p(x_{n+1}| x_n^{(k)})$. Mathematically,
\begin{equation}
TE_{Y \rightarrow X}=\sum_{i,j}p(x_{n+1},x_{n}^{(s)},y_{n}^{(t)})\log \frac{p(x_{n+1}|x_{n}^{(s)},y_{n}^{(t)})}{p(x_{n+1}|x_{n}^{(s)})}.
 \label{eq_TE}
\end{equation}
Here, $(s)$ and $(t)$ denote the number of past states of $X$ and $Y$ respectively and $n$ represents the index of the current temporal sample. If $X$ and $Y$ are independent processes, then $p(x_{n+1}| x_n^{(s)}, y_n^{(t)}) = p(x_{n+1}| x_n^{(s)})$ for all $n$, $s$, $t$ and hence $TE_{Y \rightarrow X} = 0$ bits from Eqn.~\ref{eq_TE} above. Intuitively, $TE_{Y \rightarrow X}$ attempts to capture the flow of information (in bits) from a process $Y$ to another process $X$. In general, $TE_{Y \rightarrow X} \neq TE_{X \rightarrow Y}$.  It has also been shown that TE is entirely equivalent to Granger Causality for Gaussian variables~\cite{barnett2009granger}.

We apply TE on both the multi-variate systems $\{X_1, X_2, X_3 \}$ and $\{X_1^{'}, X_2^{'}, X_3^{'} \}$. To this end, we employ the MuTE~\cite{montalto2014mute} toolbox and choose both the $binnue$ and $nnnue$ non-uniform embedding as the binning estimator for joint probability distributions. The number of bins was set to 3 (as there are only three symbols in the sequences) and the maximum past lags considered was 5. 100 surrogates with 0.05 significance level was chosen. Tables~\ref{table3} and~\ref{table4} show the results obtained for the multi-variate systems  $\{X_1, X_2, X_3 \}$ and $\{X_1^{'}, X_2^{'}, X_3^{'} \}$ respectively. All TE values are in bits. 
\begin{table}[!h]
  \caption{Transfer Entropy applied on the multi-variate system $\{X_1, X_2, X_3 \}$.}
  \label{table3}
  \centering
  \begin{tabular}{|c|c|c|}
    \hline
    Pair     & TE (binnue)    & TE (nnnue)  \\
    \hline
    $(X_1,X_2)$ & 0.2819 & 0 \\
    $(X_2,X_1)$ & 0 & 0 \\
    \hline
    $(X_1,X_3)$ & 0.2819 & 0 \\ 
    $(X_3,X_1)$ & 0 & 0 \\ 
    \hline
    $(X_2,X_3)$ & 0.2599 & 1.3195 \\
    $(X_3,X_2)$ & 0.2599 & 1.3195 \\
    \hline
  \end{tabular}
\end{table}
\begin{table}[!h]
  \caption{Transfer Entropy applied on the multi-variate system $\{X_1^{'}, X_2^{'}, X_3^{'} \}$.}
  \label{table4}
  \centering
  \begin{tabular}{|c|c|c|}
    \hline
    Pair     & TE (binnue)    & TE (nnnue)  \\
    \hline
    $(X_1^{'},X_2^{'})$ & 0.0682 & 0.9381 \\
    $(X_2^{'},X_1^{'})$ & 0.3217 & 0 \\
    \hline
    $(X_1^{'},X_3^{'})$ & 0.5737 & 0.9120 \\ 
    $(X_3^{'},X_1^{'})$ & 0.5196 & 0 \\ 
    \hline
    $(X_2^{'},X_3^{'})$ & 0 & 0 \\
    $(X_3^{'},X_2^{'})$ & 0.1819 & 0 \\
     \hline
  \end{tabular}
\end{table}
Results for TE (in Tables~\ref{table3} and \ref{table4}) show the problems in estimating causality for such a small network using infotheoretic approach. For the $\{X_1, X_2, X_3 \}$ system, the $nnnue$ setting yields correct results but fails for the system $\{X_1^{'}, X_2^{'}, X_3^{'} \}$. The $binnue$ setting does not yield meaningful results in both cases. Thus, choosing the right settings for TE is a challenging task even for small networks. To test whether TE estimates would improve with a longer data length, we doubled the lenghts. Thus, we took:
\begin{eqnarray*}
X_1 &=& [A A A A A A A A],\\
X_2 &=& [B C B C B C B C],\\
X_3 &=& [C B C B C B C B].
\end{eqnarray*}
But results did not improve at all (hence we don't include them here). Thus the problem is not just with finite length effects, but more of a fundamental nature, namely, reliably estimating probability densities in order to estimate infotheoretic quantities accurately.  
\begin{figure} [!h]
  \centering
 \includegraphics[width=1.0\linewidth]{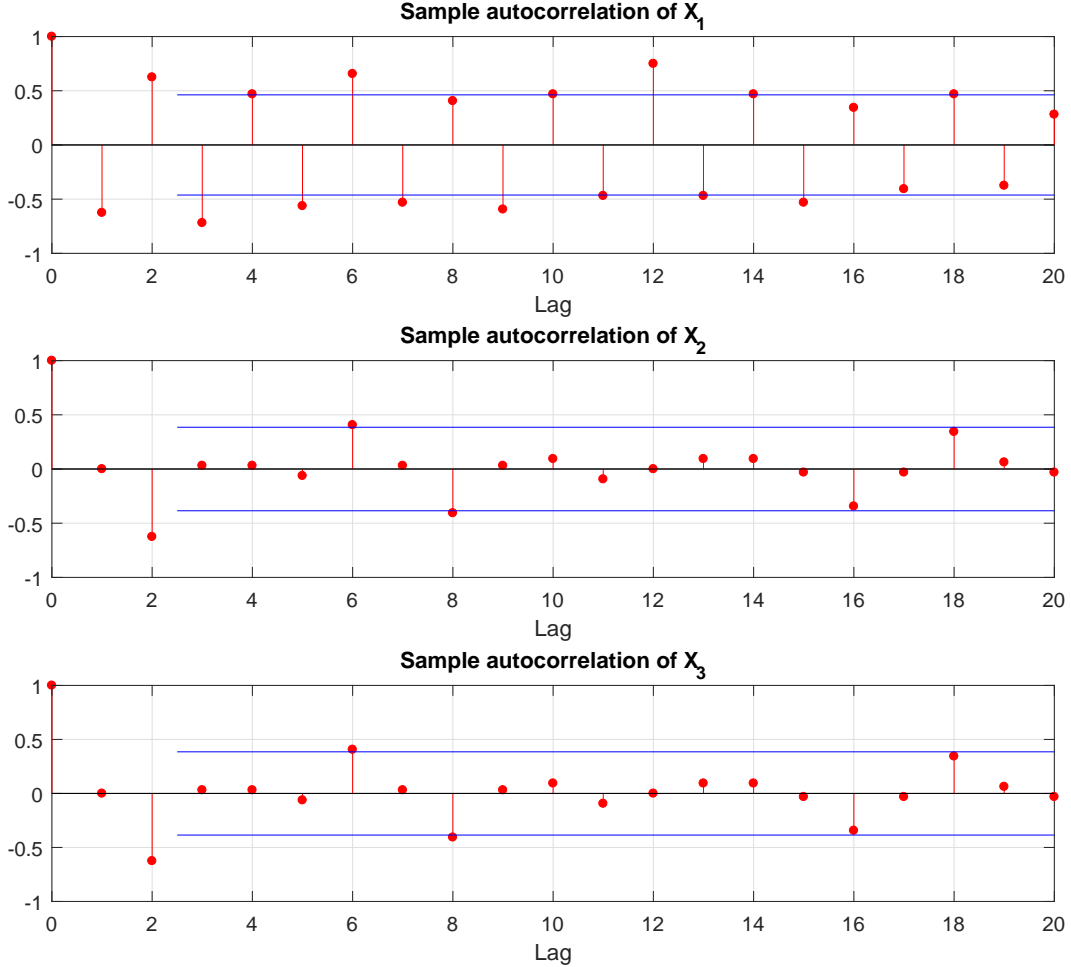}
  \caption{Sample autocorrelation function of $X_1$, $X_2$ and $X_3$ with $95\%$ confidence. A lag of 12 gives the highest autocorrelation for $X_1$. Number of lags needs to be set to a value $\geq 12$ for the $binnue$ estimator in MuTE toolbox to yield the correct causal relationships and strengths.}
  \label{fig2}
\end{figure}
\subsection{Number of lags set in TE estimation (MuTE toolbox) is critical}
For the example used in this study, one way to get the `correct' results is to use higher order estimates. In TE computation (using MuTE toolbox), this is achieved by a prudent setting of the number of maximum past lags for $binnue$ and $nnnue$ estimators. Since the firing patters $A$, $B$ and $C$ are all of length $12$, by setting the number of lags $=12$ (in MuTE) we can get the correct causal relationships and strengths of TE. With $binnue$ settings, TE($X_2 \rightarrow X_3$) $=$ TE($X_3 \rightarrow X_2$) $=$ $1.0986$ bits and TE($X^{'}_2 \rightarrow X^{'}_3$) $=$ TE($X^{'}_3 \rightarrow X^{'}_2$) $=$ $0.6931$ bits (values of TE for the remaining pairs are zero). However, there was no success for the $nnnue$ estimator as the results don't change even when the number of lags is set to 12.

Thus, in real world applications, determining the number of lags for TE is extremely critical to estimate the correct causal relationships. One way to go about determining the lags to be used is to look at autocorrelation plots. Fig.~\ref{fig2} shows the sample autocorrelation function plots for the three neurons $X_1$, $X_2$ and $X_3$.  A lag of 12 gives the highest autocorrelation for $X_1$. Number of lags needs to be set to a value $\geq 12$ for the $binnue$ estimator in MuTE toolbox to yield the correct causal relationships and strengths.

\section{Discussion on some uncomfortable aspects of information theory}

In this section, we draw attention on some uncomfortable aspects of important and central concepts of information theory that needs to be engaged by any serious researcher in causal learning.

\subsection{Conditioning reduces entropy in theory, but does it (should it) always in practice?}
A well known inequality in information theory is the following. Given two random variables $X$ and $Y$, the following is always true~\cite{cover1999elements}:
\begin{equation}
    H(X|Y) \leq H(X), 
    \label{eqn_conditioning}
\end{equation}
where $H(X|Y)$ and $H(X)$ stand for conditional entropy (of $X$ given full knowledge of $Y$) and Shannon entropy of $X$ respectively.\footnote{Assuming that these quantities exist.} Equality is achieved when $X$ and $Y$ are statistically independent. However, there could be practical scenarios where the above inequality is (or needs to be) violated, some examples are:
\begin{enumerate}
    \item In the example in section 2, knowing $X_2^{'}$ completely determines $X_3^{'}$ (since by construction they are cyclic permutations of each other) and vice versa, implying $H(X_2^{'} | X_3^{'}) = 0$ and $H(X_3^{'} | X_2^{'}) =0$ bits. However, the MI estimated from first-order statistics is zero implying that $X_2^{'}$ and $X_3^{'}$ are {\it independent}, and hence $H(X_2^{'} | X_3^{'}) = H(X_2^{'}) \neq 0$, and $H(X_3^{'} | X_2^{'}) = H(X_3^{'}) \neq 0$.   In fact, for this case, the estimated first-order Shannon entropies are $H(X_2^{'}) = H(X_3^{'}) = 1.0$ bit. Thus, we wrongly infer from first-order estimates that knowing $X_3^{'}$ is of no use in reducing the uncertainty of $X_2^{'}$ since they are (wrongly) inferred to be {\it independent}, when in fact they are completely deterministic. This is an extreme case of an himalayan blunder where {\it complete dependence} is mistaken for its opposite - namely, {\it independence}. In actual practical scenarios, it may be less disastrous (or may be not!).

    \item What about scenarios where entropies don't exist? This is very much possible if the distribution is non-stationary (for eg., varying with time) or the information source is non-ergodic.  This is probably true in many practical scenarios. Another aspect to consider is the role of noise - could be either devastating or constructive (stochastic resonance) depending on the context. 
    
    \item An important point of Shannon's theorems pertaining to source coding and channel capacity that is sometimes missed is that they assume the property of  memorylessness. Thus, many of these inequalities are invalid if the source/channel under consideration have memory. In many real-world situations, a rigorous accounting for the memory present is nontrivial, and may even be impossible. For example, if one were to ask a human to list out all of his past memories - this is an impossible task. However, in every decision making event, some specific past experience or memory of the human is brought to bear. The human may not even be aware of the presence of this memory until it is triggered in the moment of decision making.\footnote{Repressed feelings/ memories, phobias, and strong emotions are often triggered accidentally or unconsciously.}
    
    \item Is it possible for the inequality in Eq.~\ref{eqn_conditioning} to be actually reversed? The answer is {\bf Yes}, though not in the strict formalism of information theory as laid out by Shannon. If semantic information were to be considered, the above inequality can be easily violated. The possibility of \emph{confusion} in knowledge of $X$ can arise by knowing about $Y$. This is clearly true in most of human decision making scenarios in real-life. Consider the following hypothetical scenario. Tara has applied for graduate programme at several universities and she has received offers from both MIT and Stanford University. Her dream was to study at MIT and as she is about to make this decision when Stanford offers her a newly constituted `best incoming student' award (a possibility which did not exist before Tara applied). Now, she is confused.  Here, knowing {\it new} information (the award from Stanford) drastically increased Tara's uncertainty in making a decision whether to join MIT or not. 
\end{enumerate}

\subsection{Decision making by a conscious causal agent}

The above example opens  a very interesting challenge for information theoretic approach to causal reasoning. It is well understood that Shannon's notion of information or average uncertainty or degree of surprise is all about {\it making a single choice from a list of given known choices}.\footnote{Shannon was concerned mainly about transmission of a message from the sender to the receiver and not in their semantic content.} However, practical real-world decision making is rarely determinate in the {\it number} and {\it type} of choices. There are always an unknown number (and unknown kind) of unknown possibilities lurking in the background.\footnote{The ongoing pandemic is a great example of this.} In such a scenario, it is {\it impossible} to determine the sample space of the random variable as {\it new} outcomes/choices may pop up in the future (while possibly destroying some existing choices/options as well). The classical probability theoretic framework on which Shannon's information theory is based is clearly inadequate in handling such scenarios. One option is to explore the formalism of quantum theory of probability and decisions~\cite{deutsch1999quantum} which could overcome some of these limitations. 
    
From a causal learning point of view, it is important to recognize that the hallmark of a causal agent is the ability to make {\it conscious choices} and {\it decisions} exercising {\it free-will}.  In fact, a {\it conscious causal agent can  create new choices} where none existed previously.    
    
This brings us to the most elusive aspect of a {\it conscious causal agent} -- ``Consciousness''.  Until we get a grip on this most important aspect of intelligence, causal learning is a distant dream.

\section{Concluding remarks and Shannon's warning}
By means of a carefully designed illustrative example, the pitfalls of first-order estimates of infotheoretic quantities (such as MI) were demonstrated. The short data lengths forces one to use first-order estimates which leads to disastrous inferences. Data compression based measures (compression-complexity measure) seem to be more robust to finite data length effects. However,  extensive and rigorous testing of these measures (ETC, METC) needs to be performed in actual practical applications in machine learning and causal reasoning to make an informed judgement and evaluation. Causal discovery algorithms using data compression methods definitely needs a careful look.

The limitations of using the formalism of information theory is quite obvious - it fails to capture semantic aspect of information and the ability of causal agents to make decisions in an ever changing world of possibilities. A conscious causal agent can create new choices/ decisions/ meaning consciously and this is what makes causal learning possible in humans (and possibly in other animals). Consciousness is the holy grail for our understanding of intelligence, learning and causality. 

Let us conclude this with a warning made by none other than the father of information theory -- Claude E Shannon~\cite{shannon1956bandwagon}. Referring to the rapid mushrooming of applications of information theory, he wrote in~~\cite{shannon1956bandwagon}:  ``...it has perhaps been ballooned to an importance beyond its actual accomplishments.'' Also, in the same article he mentions ``...it carries at the same time an element of danger'', ``...it is certainly no panacea for the communication engineer or, {\it a fortiori}, for anyone else.'' ``It will be all too easy for our somewhat artificial prosperity to collapse overnight when it is realized that the use of a few exciting words like {\it information}, {\it entropy}, {\it redundancy}, do not solve all our problems.'' So have we been forewarned.   
\section*{Acknowledgment}
Financial support of `Cognitive Science Research Initiative' (CSRI-DST) Grant No. DST/CSRI/2017/54, `Science and Technology for Yoga and Meditation' (SATYAM-DST) Grant No.  DST/SATYAM/2017/45(G) and Tata Trusts is gratefully acknowledged. The author is thankful to Aditi Kathpalia for the help rendered on the MuTE toolbox. 

\bibliographystyle{unsrt}  
{
\small
\bibliography{references} 

\begin{thebibliography}{10}

\bibitem{shannon1948}
Claude~Elwood Shannon.
\newblock A mathematical theory of communication.
\newblock {\em The Bell system technical journal}, 27(3):379--423, 1948.

\bibitem{quinlan1986induction}
J.~Ross Quinlan.
\newblock Induction of decision trees.
\newblock {\em Machine learning}, 1(1):81--106, 1986.

\bibitem{tishby2000information}
Naftali Tishby, Fernando~C Pereira, and William Bialek.
\newblock The information bottleneck method.
\newblock {\em arXiv preprint physics/0004057}, 2000.

\bibitem{schreiber2000measuring}
Thomas Schreiber.
\newblock Measuring information transfer.
\newblock {\em Physical review letters}, 85(2):461, 2000.

\bibitem{paluvs2001synchronization}
Milan Palu{\v{s}}, Vladim{\'\i}r Kom{\'a}rek, Zbyn{\v{e}}k
  Hrn{\v{c}}{\'\i}{\v{r}}, and Katalin {\v{S}}t{\v{e}}rbov{\'a}.
\newblock Synchronization as adjustment of information rates: Detection from
  bivariate time series.
\newblock {\em Physical Review E}, 63(4):046211, 2001.

\bibitem{baghli2006model}
Mustapha Baghli.
\newblock A model-free characterization of causality.
\newblock {\em Economics Letters}, 91(3):380--388, 2006.

\bibitem{san2016information}
X~San~Liang.
\newblock Information flow and causality as rigorous notions ab initio.
\newblock {\em Physical Review E}, 94(5):052201, 2016.

\bibitem{hlavavckova2007causality}
Katerina Hlav{\'a}{\v{c}}kov{\'a}-Schindler, Milan Palu{\v{s}}, Martin
  Vejmelka, and Joydeep Bhattacharya.
\newblock Causality detection based on information-theoretic approaches in time
  series analysis.
\newblock {\em Physics Reports}, 441(1):1--46, 2007.

\bibitem{cover1999elements}
Thomas~M Cover.
\newblock {\em Elements of information theory}.
\newblock John Wiley \& Sons, 1999.

\bibitem{lzcomplexity}
Abraham Lempel and Jacob Ziv.
\newblock On the complexity of finite sequences.
\newblock {\em IEEE Transactions on information theory}, 22(1):75--81, 1976.

\bibitem{amigo2004estimating}
Jos{\'e}~M Amig{\'o}, Janusz Szczepa{\'n}ski, Elek Wajnryb, and Maria~V
  Sanchez-Vives.
\newblock Estimating the entropy rate of spike trains via lempel-ziv
  complexity.
\newblock {\em Neural Computation}, 16(4):717--736, 2004.

\bibitem{bossomaierintroduction}
T~Bossomaier, L~Barnett, M~Harr{\'e}, and JT~Lizier.
\newblock An introduction to transfer entropy: Information flow in complex
  systems. cham, switzerland: Springer international publishing; 2016.

\bibitem{barnett2009granger}
Lionel Barnett, Adam~B Barrett, and Anil~K Seth.
\newblock Granger causality and transfer entropy are equivalent for gaussian
  variables.
\newblock {\em Physical review letters}, 103(23):238701, 2009.

\bibitem{montalto2014mute}
Alessandro Montalto, Luca Faes, and Daniele Marinazzo.
\newblock Mute: a matlab toolbox to compare established and novel estimators of
  the multivariate transfer entropy.
\newblock {\em PloS one}, 9(10):e109462, 2014.

\bibitem{deutsch1999quantum}
David Deutsch.
\newblock Quantum theory of probability and decisions.
\newblock {\em Proceedings of the Royal Society of London. Series A:
  Mathematical, Physical and Engineering Sciences}, 455(1988):3129--3137, 1999.

\bibitem{shannon1956bandwagon}
Claude~E Shannon.
\newblock The bandwagon.
\newblock {\em IRE transactions on Information Theory}, 2(1):3, 1956.

\bibitem{nagaraj1}
Nithin Nagaraj, Karthi Balasubramanian, and Sutirth Dey.
\newblock A new complexity measure for time series analysis and classification.
\newblock {\em The European Physical Journal Special Topics},
  222(3-4):847--860, 2013.

\bibitem{ebeling1980grammars}
Werner Ebeling and Miguel~A Jim{\'e}nez-Monta{\~n}o.
\newblock On grammars, complexity, and information measures of biological
  macromolecules.
\newblock {\em Mathematical Biosciences}, 52(1-2):53--71, 1980.

\bibitem{benedetto2006non}
Dario Benedetto, Emanuele Caglioti, and Davide Gabrielli.
\newblock Non-sequential recursive pair substitution: some rigorous results.
\newblock {\em Journal of Statistical Mechanics: Theory and Experiment},
  2006(09):P09011, 2006.

\bibitem{grassberger2002data}
Peter Grassberger.
\newblock Data compression and entropy estimates by non-sequential recursive
  pair substitution.
\newblock {\em arXiv preprint physics/0207023}, 2002.

\bibitem{calcagnile2010non}
Lucio~M Calcagnile, Stefano Galatolo, and Giulia Menconi.
\newblock Non-sequential recursive pair substitutions and numerical entropy
  estimates in symbolic dynamical systems.
\newblock {\em Journal of nonlinear science}, 20(6):723--745, 2010.

\bibitem{benedetto2010relative}
Dario Benedetto, Emanuele Caglioti, Giampaolo Cristadoro, and Mirko
  Degli~Esposti.
\newblock Relative entropy via non-sequential recursive pair substitution.
\newblock {\em Journal of Statistical Mechanics: Theory and Experiment},
  2010(09):P09010, 2010.

\bibitem{pranay2021causal}
SY~Pranay and Nithin Nagaraj.
\newblock Causal discovery using compression-complexity measures.
\newblock {\em Journal of Biomedical Informatics}, 117:103724, 2021.

\bibitem{kathpalia2019data}
Aditi Kathpalia and Nithin Nagaraj.
\newblock Data-based intervention approach for complexity-causality measure.
\newblock {\em PeerJ Computer Science}, 5:e196, 2019.

\bibitem{virmani2019novel}
Mohit Virmani and Nithin Nagaraj.
\newblock A novel perturbation based compression complexity measure for
  networks.
\newblock {\em Heliyon}, 5(2):e01181, 2019.

\bibitem{PEERJ}
Karthi Balasubramanian and Nithin Nagaraj.
\newblock Aging and cardiovascular complexity: effect of the length of {RR}
  tachograms.
\newblock {\em PeerJ}, 4:e2755, 2016.

\bibitem{nagaraj2}
Nithin Nagaraj and Karthi Balasubramanian.
\newblock Dynamical complexity of short and noisy time series.
\newblock {\em The European Physical Journal Special Topics}, pages 1--14,
  2017.

\bibitem{nagaraj3}
Nithin Nagaraj and Karthi Balasubramanian.
\newblock Three perspectives on complexity: entropy, compression, subsymmetry.
\newblock {\em Eur. Phys. Journal Spec. Topics}, 226(15):3251--3272, 2017.

\end{thebibliography}
}
%
\appendix
\section{Appendix}


In this section, for the sake of completeness, joint entropy, conditional entropy, mutual information,  Effort-To-Compress (ETC) and Mutual ETC (METC) are described. 

\subsection{Joint entropy, conditional entropy and mutual information}
\label{MI}
Let $X$ and $Y$ be two discrete random variables with $K_1$ and $K_2$ outcomes respectively. Then, the joint entropy $H(X,Y)$ (bits/pair of symbols) is given by:
\begin{eqnarray*}
    H(X,Y) &=& -\sum_{i=1}^{K_1} \sum_{i=1}^{K_2} p_{XY}(i,j) \log_2(p_{XY}(i,j)),
\end{eqnarray*}
where $p_{XY}(\cdot)$ is the joint probability density function (assuming none of the probabilities are zeros). Conditional entropy and mutual information (MI) are defined as:
\begin{eqnarray*}
 H(X|Y) &=& H(X,Y) - H(Y), \\
    H(Y|X) &=& H(X,Y) - H(X), \\
       MI(X,Y) &=& H(X) + H(Y) - H(X,Y).
\end{eqnarray*}

\subsection{Effort-to-Compress (ETC) and Mutual ETC (METC)}
\label{meth_ETC}
Effort-To-Compress or ETC~\cite{nagaraj1} is a ``compression-complexity measure'' that is derived from a lossless data compression algorithm known as Non-Sequential Recursive Pair Substitution (NSRPS).
algorithm~\cite{ebeling1980grammars}. For an input sequence (of symbols),  ETC proceeds by parsing the sequence in iterations. In the first iteration, the most frequently occurring pair of symbols in the sequence is replaced by a new symbol (not present in the original sequence). This leads to a smaller derived sequence. Subsequent iterations proceed in the same manner - each time creating a derived sequence shorter than the previous. The algorithm halts when the entire sequence has only one unique symbol. The value of ETC is defined as ``the
number of iterations needed to reduce the input sequence into a constant sequence''.
As an example, consider the input sequence `-1 -1 0 1 -1 -1  1 0'. This gets transformed as follows: `-1 -1 0 1 -1 -1  1 0' $\rightarrow$ `2 0 1 2 1 0' $\rightarrow$ `3 1 2 1 0' $\rightarrow$ `4 2 1 0' $\rightarrow$ `5 1 0' $\rightarrow$ `6 0' $\rightarrow$ '7'.  The value of ETC is 6. This can be normalized by dividing by $L-1$ where $L$ is the length of the original sequence. In this case, the normalized value of ETC will be $6/8$.  A higher value of ETC indicates higher complexity of the original sequence. 

In order to compute Mutual ETC (METC) between two sequences, the above ETC algorithm (which is for 1D sequences) is first extended for 2D sequences (ETC$_{2D}$ or joint ETC). The extension is straightforward and hence omitted here (basically, consider pairs of 2D vectors of symbols instead of individual symbols). Mutual ETC or METC between two sequences $X$ and $Y$ is computed as:
\begin{equation}
    METC(X,Y) = ETC(X) + ETC(Y) - ETC_{2D}(X,Y), 
\end{equation}
where $ETC(\cdot)$ refers to the 1D version (described in the paragraph above with an example).

Papers related to rigorous mathematical  properties of NSRPS, lossless data compression properties, and entropy estimates using NSRPS can be found in~\cite{benedetto2006non, grassberger2002data, calcagnile2010non, benedetto2010relative}. Compression-complexity measures find applications in causal discovery~\cite{pranay2021causal}, causality testing~\cite{kathpalia2019data}, dynamical complexity of brain networks~\cite{virmani2019novel} and cardiovascular complexity~\cite{PEERJ}. Specifically, they have been found to work well with short time series~\cite{nagaraj2, nagaraj3}.

\end{document}